\documentclass[aps, prl, twocolumn, superscriptaddress, showpacs]{revtex4}
\usepackage{graphicx}
\usepackage{amsmath,amssymb}
\begin{document}

\title{ Engineering the Kondo and Fano effects in double quantum dots}
\author{Tie-Feng Fang}
\affiliation{Center for Interdisciplinary Studies, Lanzhou University, Lanzhou 730000, China}
\author{Hong-Gang Luo}
\affiliation{Center for Interdisciplinary Studies, Lanzhou University, Lanzhou 730000, China} \affiliation{Institute of Theoretical Physics, Chinese Academy of Sciences, Beijing 100080, China}
\begin{abstract}
We demonstrate delicate control over the Kondo effect and its interplay with quantum interference in an Aharonov-Bohm interferometer containing one Kondo dot and one noninteracting dot. It is shown that the Kondo resonance undergoes a dramatic evolution as the interdot tunnel coupling progressively increases. A novel triple Kondo splitting occurs from the interference between constant and Lorentzian conduction bands that cooperate in forming the Kondo singlet. The device also manifests a highly controllable Fano-Kondo effect in coherent electronic transport, and can be tuned to a regime where the coupled dots behave as decoupled dots.
\end{abstract}
\maketitle

The Kondo effect is a well-known strongly correlated phenomenon in condensed matter physics \cite{Hewson}. It describes a collective state known as Kondo resonance around the Fermi level due to the screening of an impurity spin by conduction electrons. Since its first experimental observation in quantum dots (QDs) \cite{Goldhaber-Gordon}, unprecedented control over this many-body effect has been systematically achieved by exploiting the tunable physical characteristics of QDs \cite{QD} and by fabricating complicated QD devices, such as coupled multiple dots \cite{Chang} or dots embedded in an Aharonov-Bohm (AB) geometry \cite{ABexp,ABthe}. Interestingly, the Kondo effect in coherent transport through multipath geometries may get entangled with the Fano resonance \cite{Miroshnichenko}, an interference effect between resonant and nonresonant channels.

Recently, double QDs consisting of one Kondo dot and one noninteracting dot have been suggested as a promising platform to achieve novel Kondo physics \cite{Silva2009,Silva2006,Vaugier}. For example, a crossover from a non-Kondo dip to a conventional Kondo peak is realized when one adjusts the magnetic flux in the AB interferometer \cite{Silva2009}, which equivalently manipulates the effective pseudogapped band. And a double Kondo splitting develops in the side-dot geometry with increasing interdot coupling \cite{Silva2006} or decreasing dot-lead coupling \cite{Vaugier}. Although these phenomena are intriguing they have been obtained in two distinct configurations by tuning different parameters. This motivates us to explore a configuration, in which these features as well as some possibly novel ones can be captured by continuously tuning only one parameter. This is also of experimental interest since as few control parameters as possible is more accessible in experiments.

In this letter, we propose such a device, i.e., a fully connected AB interferometer where the Kondo dot (dot 1) and the noninteracting dot (dot 2) are tunnel coupled to left ($L$) and right ($R$) leads as well as to each other (see Fig.\,1). The direct interdot tunneling divides the geometry into two hopping loops, in which two applied magnetic fluxes can independently control the interference condition. Under suitable conditions, we show that on progressive increase of this interdot coupling, the local density of states (DOS) on the Kondo dot dramatically evolves from a non-Kondo dip to a sharp Kondo peak, followed by a triple splitting of the peak, and eventually the triple splitting reduces to a double splitting. Compared with previous studies \cite{Silva2009,Silva2006,Vaugier} we emphasize that these features are continuously produced in a single geometry by only tuning the interdot tunneling (via a gate voltage in experiments). In particular, the novel triple Kondo splitting not reported previously is ascribed to the interference between the constant and Lorentzian components of the effective conduction band that have nearly equal contribution to screen the impurity spin. Moreover, our device gives even deeper insight into the interplay of many-body correlations and quantum interference, exhibiting an unambiguous Fano-Kondo effect with a highly tunable Fano parameter in the linear conductance, and accessing a striking regime where the coupled dots exhibit decoupled local and transport properties.

The device depicted in Fig.\,1 can be modeled by an Anderson-like Hamiltonian $H=H_0+H_D+H_V$, where $H_0=\sum_{ k, \sigma, \alpha }\varepsilon_{ k }C_{ k \sigma \alpha }^{ \dagger }C_{ k \sigma \alpha }$ ($\alpha=L,R$) describes the leads with constant DOS $\rho_0$, $H_D=\sum_{ m, \sigma }\varepsilon_{ m }d_{ m \sigma }^{ \dagger }d_{ m \sigma }+Ud^\dagger_{ 1 \uparrow }d_{1 \uparrow }d^\dagger_{ 1 \downarrow }d_{ 1 \downarrow }$ ($m=1,2$) models the isolated dots, and $H_V=\sum_{ \sigma }Jd_{ 1 \sigma }^{ \dagger }d_{ 2 \sigma }+\sum_{ m, k, \sigma, \alpha }V_{ m \alpha }d_{ m \sigma }^{ \dagger }C_{ k \sigma \alpha }+\textrm{H.c.}$ is the dot-dot ($J\equiv|J|e^{i\phi_J}$) and dot-lead ($V_{m\alpha}\equiv|V_{m\alpha}|e^{i\phi_{m\alpha}}$) couplings. Here $d^\dagger_{m\sigma}$ ($C^\dagger_{k\sigma\alpha}$) creates a spin-$\sigma$ electron of energy $\varepsilon_m$ ($\varepsilon_k$) in dot $m$ (the $\alpha$ lead), and the on-site Coulomb repulsion $U$ is restricted within dot $1$. The intrinsic linewidth of the dot levels due to tunnel coupling to the leads is $\Gamma_m=\sum_\alpha\Gamma_{m\alpha}$ with $\Gamma_{ m \alpha }=\pi\rho_0|V_{ m \alpha }|^2$. The magnetic flux $\Phi_\alpha$ penetrating the $\alpha$ hopping loop introduces an AB phase $\phi_\alpha=2\pi\Phi_\alpha/\Phi_0$, with $\Phi_0$ being the flux quantum, so that $\phi_{ 1 \alpha }-\phi_{ 2 \alpha }-\phi_J=(\delta_{ \alpha L }-\delta_{ \alpha R })\phi_\alpha$.

\begin{figure}[ht]
\begin{center}
\includegraphics[width=0.7\columnwidth]{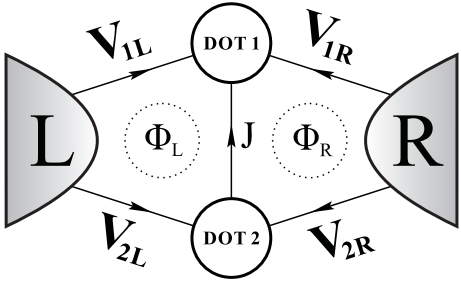}
\caption{Double-QD AB interferometer with direct interdot tunneling and two independent magnetic fluxes.}
\end{center}
\end{figure}

This model describes an Anderson impurity (dot $1$) embedded in a complex but noninteracting network (dot $2$ and the leads). Its local and transport properties are completely determined by the dot-$1$ retarded Green's function (GF), $G_1( \varepsilon )$, which we calculate using the equation of motion (EOM) approach \cite{Lacroix,Luo,OEW}. The EOM for $G_1( \varepsilon )$ involves a high-order GF whose EOM produces in turn more higher-order ones. We truncate this hierarchy by i) the Lacroix's decoupling procedure \cite{Lacroix,Luo} generalized to the case of arbitrary networks \cite{OEW} and ii) neglecting higher-order GFs that contribute only at order $1/U^2$. By this two-step scheme \cite{note1}, in the wide-band limit and taking $\Gamma_{mL}=\Gamma_{mR}$, the dot-$1$ GF is given by
\begin{equation}
G_1(\varepsilon)=\frac{\widetilde{\Xi}(\varepsilon)-U[1-n_{1}-P(\varepsilon)]}
{\Xi(\varepsilon)\widetilde{\Xi}(\varepsilon)-U[\Xi(\varepsilon) +\Sigma_0(\varepsilon)P(\varepsilon)-Q(\varepsilon)]},
\end{equation}
where $\Xi(\varepsilon)=\varepsilon-\varepsilon_1-\Sigma_0(\varepsilon)$, $\widetilde\Xi(\varepsilon)=\Xi(\varepsilon)-\Sigma_0(\varepsilon)$, and
\begin{eqnarray}
n_1&=&-\frac{1}{\pi}\int\textrm{d}\varepsilon\,f(\varepsilon)\textrm{Im}[G_1(\varepsilon)],\\
P\left( \varepsilon \right)&=&\frac{1}{\pi }\int \textrm{d}\varepsilon ^{\prime }\,
\frac{f\left( \varepsilon ^{\prime }\right) }{\varepsilon -\varepsilon
^{\prime }}\big\lbrace\Sigma _{0}\left( \varepsilon \right)
\textrm{Im}\left[ G_{1}\left( \varepsilon ^{\prime }\right) \right]\nonumber\\
&&-\textrm{Im}\left[ \Sigma _{0}\left( \varepsilon ^{\prime }\right) G_{1}\left(
\varepsilon ^{\prime }\right) \right]\big\rbrace,\\
Q\left( \varepsilon \right)&=&\frac{1}{\pi }\int \textrm{d}\varepsilon ^{\prime }\,
\frac{f\left( \varepsilon ^{\prime }\right) }{\varepsilon -\varepsilon
^{\prime }}\big\lbrace \Sigma _{0}\left( \varepsilon \right) \textrm{Im}\left[
\Sigma _{0}\left( \varepsilon ^{\prime }\right) G_{1}\left( \varepsilon
^{\prime }\right) \right]\nonumber\\
&&-\textrm{Im}\left[ \Sigma _{0}\left( \varepsilon
^{\prime }\right) +\Sigma _{0}^{2}\left( \varepsilon ^{\prime }\right)
G_{1}\left( \varepsilon ^{\prime }\right) \right]\big\rbrace,\\
\Sigma_0(\varepsilon)&=&-i\Gamma_1+\big[|J|^2-\Gamma_1\Gamma_2\cos^2\frac{ \phi_L + \phi_R }{2}\nonumber\\
&&-(i/8)|J|\sqrt{ \Gamma_1 \Gamma_2 }(\cos\phi_L+\cos\phi_R)\big] g_2(\varepsilon).
\end{eqnarray}
Here $\Sigma_0(\varepsilon)$ is the self-energy of $G_1(\varepsilon)$ for $U=0$, $f(\varepsilon)$ represents the Fermi distribution function, and $g_2(\varepsilon)=1/( \varepsilon - \varepsilon_2 + i\Gamma_2 )$ is the dot-$2$ GF in the absence of dot $1$. Eqs.\,(1)-(5) can be solved self-consistently. We emphasize that Eqs.\,(1)-(4) are quite universal and applicable to an interacting dot embedded in general complex networks whose geometrical details are incorporated in $\Sigma_0(\varepsilon)$. Our system is thus equivalent to an Anderson impurity with a structured hybridization function $\Delta(\varepsilon)\equiv-\textrm{Im}\Sigma_0(\varepsilon)$ which, in general, has an asymmetric peak-dip Fano profile, and effectively renormalizes the conduction DOS seen by the impurity, i.e., $\rho_0\to\rho^\prime_0(\varepsilon)\propto\Delta(\varepsilon)$.

To achieve the most fascinating properties, we restrict ourself in this work to a particular phase condition $\phi_L-\phi_R=(2l+1)\pi$, $l=0,\pm1,\pm2,\cdots$. Therefore Eq.\,(5) reduces to $\Sigma_0(\varepsilon)=-i\Gamma_1+\Omega g_2(\varepsilon)$ and thus $\Delta(\varepsilon)=\Gamma_1-\Omega\textrm{Im}[g_2(\varepsilon)]$, where $\Omega=|J|^2-J^2_0\cos^2(\phi/2)$ with $J_0=\sqrt{\Gamma_1\Gamma_2}$ and $\phi=\sum_\alpha\phi_\alpha$ being the total AB phase. In this case, by varying the interdot tunneling $|J|$, $\Delta(\varepsilon)$ around $\varepsilon=\varepsilon_2$ can continuously evolve from a Lorentzian dip to a Lorentzian peak of width $\Gamma_2$. In the numerical results presented below, we fix $U=12\Gamma_1=24\Gamma_2=0.12D$ with the half bandwith $D=1$. The dot levels $\varepsilon_m$ are measured from the Fermi level $\varepsilon_F=0$ which lies at the middle of the band.

\begin{figure}[ht]
\begin{center}
\includegraphics[width=1.0\columnwidth]{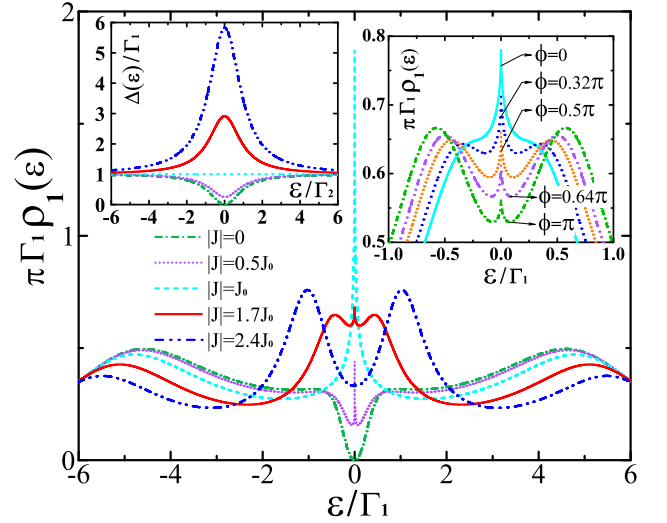}
\caption{Variation of the dot-$1$ DOS $\rho_1(\varepsilon)$ with the interdot tunneling $|J|$ at AB phase $\phi=0$. Left inset: Corresponding evolution of the hybridization function. Right inset: Triple Kondo splitting in $\rho_1(\varepsilon)$ for different AB phases at $|J|=1.56J_0$. Other parameters used are $\varepsilon_1=-U/2$, $\varepsilon_2=0$ and the temperature $T=10^{-6}\Gamma_1$.}
\end{center}
\end{figure}

Figure 2 presents the dot-$1$ DOS $\rho_1(\varepsilon)\equiv-\frac{2}{\pi}\textrm{Im}G_1(\varepsilon)$ at the particle-hole symmetric point for $\phi=0$, accompanied by the corresponding hybridization function $\Delta(\varepsilon)$ in the left inset. Upon progressively increasing the interdot tunneling $|J|$, while the broad Hubbard bands centered near $\varepsilon_1$ and $\varepsilon_1+U$ are monotonically suppressed and departed from the Fermi level, the local DOS around the Fermi level undergoes a dramatic evolution. It first features a non-Kondo dip at $|J|=0$ for which one has exactly $\Delta(\varepsilon_F)=0$, signaling a complete suppression of the Kondo effect by a pseudogap forming in the effective conduction band. The dip converts to a conventional Kondo resonance when $|J|$ approaches to a special value where $\Delta(\varepsilon)=\Gamma_1$ and thus the usual Anderson model is recovered. As $|J|$ further increases, the resonance splits into three peaks and eventually manifests only two Kondo peaks by vanishing the central peak. In general, the growing spectral weight of the Kondo resonance indicates the enhanced Kondo effect resulting from the increasing DOS near $\varepsilon_F$ of the conduction electrons which contribute to screen the excess spin in dot $1$. In particular, the splitting of the Kondo resonance implies complicated interference processes that deserve a careful analysis.

In the side-dot geometry \cite{Silva2006,Vaugier}, the Kondo resonance on the correlated dot suffers a double splitting due to its interference with the single-particle resonance on the other dot. Considering our noninteracting DOS $\rho_1^0(\varepsilon)\equiv-\frac{2}{\pi}\textrm{Im}[\Xi(\varepsilon)]^{-1}$ for $\varepsilon_1=\varepsilon_2=0$, it exhibits one (two) peak(s) for weak (strong) interdot tunneling but by no means the three-peak structure. Therefore the picture of interference between the resonances on the two dots is insufficient to interpret the triple Kondo splitting predicted here. Instead of a single Lorentzian conduction DOS in the side-dot configuration \cite{Silva2006,Vaugier}, we note that in our geometry, the effective DOS of conduction electrons [$\propto\Delta(\varepsilon)$] consists of a constant part ($\propto\Gamma_1$) from the leads and a Lorentzian part ($\propto-\Omega\textrm{Im}[g_2(\varepsilon)]$) from dot 2. The system maps thus into an Anderson impurity coupled to two reservoirs with constant and Lorentzian DOS, respectively. Unlike the two-channel Kondo effect \cite{Potok} where two independent reservoirs compete in forming the Kondo singlet, our two reservoirs actually cooperate to screen the localized spin in dot 1 because they allow exchange of electrons. As a result of this cooperation, the triple Kondo splitting occurs in the intermediate regime $|J|\sim\Gamma_1$ where the two conduction bands contribute equally. Equivalently, the triple splitting can also be viewed as a nontrivial consequence of the interference between the screening conduction electrons from the leads and dot 2. These electrons are spatially separated, having different phase shifts due to the magnetic fluxes penetrating the geometry. Therefore the AB effect can control the interference condition and thus significantly alter the lineshape of the triple splitting by transferring the spectral weight among the three peaks, as shown in the right inset of Fig.\,2. For large $|J|$, the interdot process (responsible for the Lorentzian conduction DOS) dominates the Kondo physics, similar two-peak behavior as in the side-dot geometry \cite{Silva2006,Vaugier} is thus recovered.

We now turn to the coherent electronic transport through the system. In linear response regime and at zero temperature, only single-electron elastic processes are allowed. The conductance is thus $G=\frac{2e^2}{h}|\mathcal{T}(\varepsilon_F)|^2$ from the Laudauer-B\"uttiker formula \cite{Meir}, where $\mathcal{T}(\varepsilon_F)$ is the transmission amplitude from left to right for electrons with energy $\varepsilon_F$. $\mathcal{T}(\varepsilon_F)$ can be expressed in terms of the dot-$1$ GF as $\mathcal{T}(\varepsilon_F)=e^{i(\phi_{2R}-\phi_{2L})}\big[\Gamma_2g_2(\varepsilon_F)+\lambda G_1(\varepsilon_F)\big]$
with $\lambda=\Gamma_1e^{-i(\phi_L+\phi_R)}-g_2(\varepsilon_F) \big[i\Gamma_1 \Gamma_2e^{-i(\phi_L+\phi_R)} - \Gamma_2\Sigma_0( \varepsilon_F) \big]$. For $|J|=0$, $\varepsilon_2=\varepsilon_F$, and $\phi=2l\pi$, so that $\Delta(\varepsilon_F)=0$, one has $G_1(\varepsilon_F)=0$, therefore $G=2e^2/h$ is irrelevant to $\varepsilon_1$. Otherwise, Fermi-liquid theory \cite{Hewson} yields the unitary condition $\textrm{Im}\Sigma(\varepsilon_F)=-\Delta(\varepsilon_F)$ and the Friedel sum rule $[\varepsilon_1+\textrm{Re}\Sigma(\varepsilon_F)]/\Delta(\varepsilon_F)=\cot(\pi\widetilde n_1)\equiv\tilde\varepsilon$, where $\Sigma(\varepsilon)$ is the self-energy of $G_1(\varepsilon)$ for arbitrary $U$ and $\widetilde n_1=n_1+\frac{1}{\pi}\textrm{Im}\int^{ \varepsilon_F}_{ -\infty}\textrm{ d}\varepsilon\,\frac{ \partial\Sigma_0( \varepsilon)}{ \partial\varepsilon}G_1( \varepsilon)$ is the displaced electron number per spin introduced by dot $1$. With these Fermi-liquid relations, the linear conductance is exactly expressed by the Fano formula \cite{Miroshnichenko}
\begin{equation}
G=\frac{2e^2}{h}T_2\frac{\vert\tilde\varepsilon_1+q\vert^2}{\tilde\varepsilon^2_1+1},
\end{equation}
where $T_2\equiv\Gamma_2^2/(\varepsilon_2^2+\Gamma^2_2)$ represents the transmission probability through dot $2$ in the absence of dot $1$ and the Fano parameter $q\equiv\mu\varepsilon_2/\Gamma_2+i\nu\sqrt{(1-T_2)/T_2}\sin\phi$ with $\mu\equiv1-\nu+\nu\cos\phi$ and $\nu\equiv\Gamma_1/\Delta(\varepsilon_F)$. Noteworthily, $q$ is a complex number due to the broken time reversal symmetry in the presence of the magnetic fields. The Fano effect described by Eq.\,(6) produces asymmetric lineshapes in $G$ as a function of $\varepsilon_1$, resulting from the interference between resonant transmission through dot $1$ and dot $2$. Here dot $2$ with its constant transmission amplitude serves as the continuum since $\varepsilon_2$ is fixed.

The linear conductance $G$ is calculated in Fig.\,3 as a function of the renormalized dot-$1$ level $Vg\equiv2\varepsilon_1+U$ for different values of $|J|$ at $\phi=0$. As $\varepsilon_2$  shifts across the Fermi level, generically, we find asymmetric Fano lineshapes of peaks and dips around $V_g=\pm U$ with reduced Kondo plateau at $V_g=0$ [Figs.\,3(b), 3(c), 3(d), 3(f), 3(g), and 3(h)]. However, transport through dot $2$ is blocked by placing the resonance on dot $2$ far away from the Fermi level, giving rise to the usual Kondo plateau of unitary transmission through dot $1$ [Figs.\,3(a) and 3(i)]. By contrast, when dot $2$ is exactly in resonance with the leads, the Kondo plateau falls into a complete Kondo valley [Fig.\,3(e)] due to destructive interference processes for transmission through the two dots. We emphasize that these characteristic lineshapes of Fano-Kondo effect are completely governed by the Fano parameter which, for $\phi=0$, is $q=\varepsilon_2/\Gamma_2$ independent of the interdot tunneling $|J|$. As a consequence, upon the variation of $|J|$, the residual $|J|$-dependence of $\tilde \varepsilon_1$ enables only a tiny tuning of the peaks, dips, and plateaus while the essential Fano-Kondo profile remains robust [Figs.\,3(b)-3(h)]. Furthermore, the conductance is even totally irrelevant to the interdot tunneling for $|\varepsilon_2|\gg\Gamma_2$ where dot $2$ is effectively absent from the transmission [Figs.\,3(a) and 3(i)].

\begin{figure}[ht]
\begin{center}
\includegraphics[width=0.9\columnwidth]{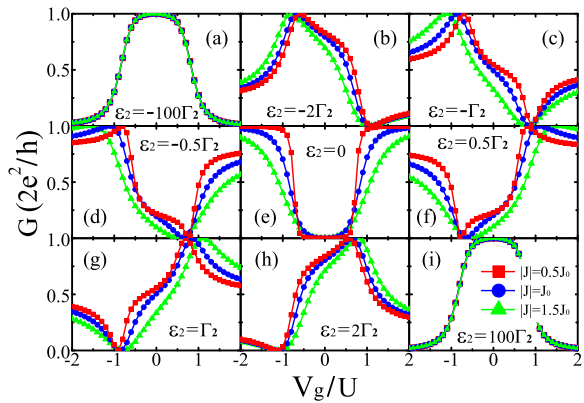}
\caption{Linear conductance $G$ as a function of the renormalized dot-$1$ level $Vg\equiv2\varepsilon_1+U$ for different values of $\varepsilon_2$ and $|J|$ at $\phi=0$. The resulting Fano parameter is $q=-100,-2,-1,-0.5,0,0.5,1,2,100$ from (a) to (i).}
\end{center}
\end{figure}

To promote the tunability of this Fano-Kondo effect upon the interdot tunneling, a nonzero total AB phase is required such that $q$ is $|J|$-dependent and $\textrm{Im}\,q$ comes into play. The case $\varepsilon_2=\Gamma_2$, which gives a bare transmission probability $T_2=0.5$ through dot $2$, yields the most striking properties shown in Fig.\,4. Here, the nature of the interference and hence the Fano parameter change significantly as the interdot tunneling or the AB phase varies, giving rise to dramatic evolutions of the lineshape. While the sign of $\textrm{Im}\,q$ is irrelevant, the lineshape of a dip to the left (right) of a peak is due to positive (negative) $\textrm{Re}\,q$. Interestingly, Fig.\,4 demonstrates a featureless lineshape at $|J|=J_0/\sqrt{2}$ and $\phi=0.5\pi$ where one has exactly $q=i$ and thus $G=e^2/h$, meaning that the linear conductance of the device is determined by transport through dot 2 as if dot 1 is absent. In this regime, the nontrivial interplay of many-body correlations and quantum interference effectively cancels the coupling between the two dots. This can also be apparently seen in the local properties. Since $\Delta(\varepsilon)=\Gamma_1$ for $|J|=J_0/\sqrt{2}$ and $\phi=0.5\pi$, the Lorentzian component of the conduction band from dot 2 is completely suppressed. Accordingly, the dot-1 DOS will exhibit the usual Kondo resonance (like the dashed line in Fig.\,2) induced by resonant spin-flip scattering of electrons only from the lead band as if dot 2 is decoupled from the system.

\begin{figure}[ht]
\begin{center}
\includegraphics[width=0.9\columnwidth]{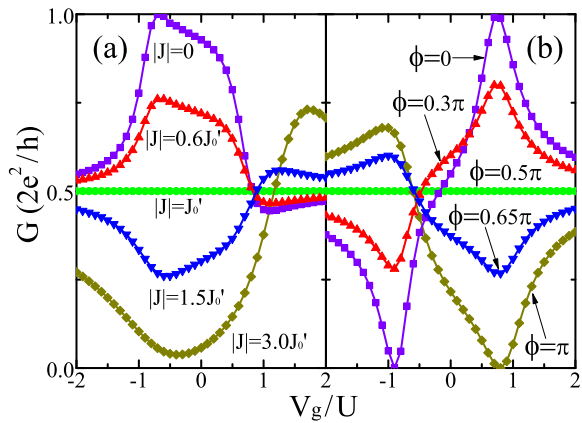}
\caption{Dependence of the Fano-Kondo lineshape on (a) the interdot tunneling at $\phi=0.5\pi$ and (b) the AB phase at $|J|=J_0/\sqrt{2}\equiv J'_0$ for $\varepsilon_2=\Gamma_2$. }
\end{center}
\end{figure}

In summary, we have studied the Kondo and Fano physics in a double-QD AB interferometer and demonstrated a dramatic manipulation of the Kondo resonance on one dot by varying its coupling to the other dot. It is further shown that the nontrivial interplay of correlations and interference manifests as an unambiguous Fano-Kondo effect in transport properties and even surprisingly decouples the two dots in a certain parameter regime. These results may promote the flexibilities for engineering QD-based devices with newly tailored properties.

Support from the NSFC and the Program for NCET of China is acknowledged.

\end{document}